\documentclass[journal,10pt,doublecolumn]{IEEEtran}
\usepackage{geometry}
\usepackage{graphicx}
\usepackage{amsfonts}
\usepackage{mathrsfs}
\usepackage{amssymb,amsmath}
\usepackage{bm}
\usepackage{algorithm}
\usepackage{algorithmic}
\usepackage{theorem}
\usepackage{array,color}
\usepackage{cite}
\usepackage{stfloats}
\usepackage{subfigure}
\usepackage{array}
\usepackage{multirow}
\usepackage{bbding}
\usepackage{epstopdf}
\newcommand{\tb}{\mathbf}

\newlength{\messagetablewidth}
\newcounter{saveequationcntr}%

{\begin{table}%
\setcounter{saveequationcntr}{\value{equation}}%
\setcounter{equation}{0}%
}%
{\setcounter{equation}{\value{saveequationcntr}}%
\end{table}}
\IEEEoverridecommandlockouts

\begin{document}
\title{A Simple Variational Bayes Detector for Orthogonal Time Frequency Space (OTFS) Modulation}
\author{Weijie Yuan, Zhiqiang Wei,~\IEEEmembership{Member,~IEEE,} Jinhong Yuan,~\IEEEmembership{Fellow,~IEEE,} and Derrick Wing Kwan Ng,~\IEEEmembership{Senior Member,~IEEE\vspace{-5mm}}
\thanks{The authors are with the School of Electrical Engineering and Telecommunications, University of New South Wales, NSW 2052, Australia.

{
 }

{}

{}
}
}

\maketitle
\begin{abstract}
The emerging orthogonal time frequency space (OTFS) modulation technique has shown its superiority to the current orthogonal frequency division multiplexing (OFDM) scheme, in terms of its capabilities of exploiting full time-frequency diversity and coping with channel dynamics.
The optimal maximum \emph{a posteriori} (MAP) detection is capable of eliminating the negative impacts of the inter-symbol interference in the delay-Doppler (DD) domain at the expense of a prohibitively high complexity.
To reduce the receiver complexity for OTFS scheme, this paper proposes a variational Bayes (VB) approach as an approximation of the optimal MAP detection.
Compared to the widely used message passing algorithm, we prove that the proposed iterative algorithm is guaranteed to converge to the global optimum of the approximated MAP detector regardless the resulting factor graph is loopy or not.
%
%
Simulation results validate the fast convergence of the proposed VB receiver and also show its promising performance gain compared to the conventional message passing algorithm.
\end{abstract}

\begin{IEEEkeywords}
Orthogonal time frequency space, receiver design, variational Bayes, low complexity
\end{IEEEkeywords}
\section{Introduction}
Future wireless communication applications require reliable data transmission in high mobility environments, such as vehicular networks and high speed railway\cite{wong2017key}.
When facing a fast time-varying channel, the widely adopted orthogonal frequency division multiplexing (OFDM) modulation in currently cellular networks suffers from a dramatic performance degradation due to the inter-carrier interference imposed by Doppler shift \cite{raviteja2018practical}.

Recently, a newly developed modulation scheme, namely, orthogonal time frequency space (OTFS) has attracted significant attention since it is robust in doubly selective channels \cite{Hadani2017orthogonal}.
The basic model of OTFS modulation/demodulation was devised in \cite{RavitejaOTFS}.
In \cite{FarhangCPOFDM} and \cite{RezazadehReyhaniCPOFDM}, the discrete time formulation of OTFS is developed and it was shown that OTFS can be implemented via appending a pre- and post-processing to conventional OFDM schemes.
In particular, OTFS modulator multiplexes information symbols in the delay-Doppler (DD) domain.
Based on a two-dimensional inverse symplectic finite Fourier transform (ISFFT), each symbol in the DD domain is spread over all lattices in the time-frequency (TF) domain.
Consequently, full time and frequency diversity can be realized and exploited, which leads to a further performance gain compared to conventional OFDM schemes in time-varying channels\cite{surabhi2019diversity}.
Moreover, when the number of reflectors/scatters is limited during propagation, employing DD domain data multiplexing provides a sparse channel representation, enabling a better channel estimation performance and a low-complexity detector.

%
In practice, the time and frequency diversity can be fully exploited with the optimal maximum \emph{a posteriori} (MAP) detection.
Yet, the complexity of the MAP detection increases exponentially with the block size of each OTFS frame.
%
%
As a compromise approach, based on the factor graph and message passing framework \cite{kschischang2001factor}, the authors in \cite{raviteja2018low} developed a low-complexity receiver applying Gaussian approximation to capture the impact of the inter-symbol interference.
Nevertheless, message passing (MP) receivers will converge to a locally optimal point in a loopy factor graph, which is very likely to happen for OTFS in a multipath scenario\cite{ihler2005loopy}.
Therefore, it is highly desirable to develop a low-complexity receiver with fast convergence for OTFS systems.

Motivated by the aforementioned discussions, in this paper, a low-complexity detector for OTFS systems is designed.
We commence from the optimal MAP detection and derive an approximation of the \emph{a posteriori} distribution exploiting Kullback-Leibler (KL) divergence to reduce the detection complexity.
Then, a variational Bayes (VB) approach is used to maximize the evidence lower bound (ELBO) iteratively, which yields a marginal distribution for each symbol and thus results in a low-complexity point-wise MAP detection.
Compared to the conventional MP receivers, the advantages of the proposed variational Bayes approach are two-fold.
Firstly, by choosing the distribution family for the VB method appropriately, the ELBO maximization problem is strictly convex such that the globally optimal solution can be obtained efficiently, leading to a convergence guaranteed receiver.
Secondly, the VB approach only requires a substantially lower complexity than that of the message passing algorithms.
Furthermore, we demonstrate the rapid convergence and the better detection performance compared to MP receivers through simulations, particularly in a practical multipath propagation environment.

\emph{Notations:} We use a boldface lowercase letter and boldface capital letter to denote a vector and a matrix, respectively.
The superscript $\textrm{T}$, $*$, {and} $\textrm{H}$ denote the transpose, conjugate, and {the} Hermittian operations{, respectively};
$\propto$ represents both sides of the equation are multiplicatively connected to a constant; $|\cdot|$ denotes the modulus of a complex number or the cardinality of a set; $[\cdot]_N$ denotes the modulo operation with divisor $N$; $\|\cdot\|$ denotes the $\ell^2$ norm; $\mathcal{R}\{x\}$ denotes the real part of complex number $x$; $\mathbb{E}_{p}[x]$ denotes the expectation of $x$ with respect to distribution $p$; $\partial$ denotes the partial derivative operator; the big-$O$ notation $\mathcal{O}$ is an asymptotic notation describes the order of complexity.
The circularly symmetric complex Gaussian distribution with mean $\mu$ and variance $\sigma^2$ is denoted by ${\cal CN}(\mu,\sigma^2)$.

\section{System Model}
The OTFS modulation and demodulation are realized by employing two dimensional pre- and post-processing.
We focus on a point-to-point scenario.
In particular, the transmitted coded bits $\tb{c}$ are mapped to data symbols $d_{k,l}\in \chi$ in the DD domain with constellation set $\chi$, where $k\in [0,N-1]$ and $l\in[0,M-1]$ denotes the index of Doppler shift and delay, respectively.
Integers $N$ and $M$ denote the numbers of available time slots and subcarriers, respectively.
The symbols $\{d_{k,l}\}$ are transformed to time-frequency (TF) domain symbols $X_{n,m}$ through ISFFT, given by\cite{RavitejaOTFS}
\begin{align}
\tb{X}= \tb{F}_N^{\textrm{H}}\tb{D}\tb{F}_M,
\end{align}
where $\tb{X}$ and $\tb{D}$ are matrices with elements $X_{n,m}$ and $d_{k,l}$ and $\tb{F}_N$ denotes the normalized discrete Fourier transform (DFT) matrix of dimension $N$.
Then, the TF domain symbols $\tb{X}$ are converted to a continuous transmitted signal $s(t)$, expressed as
\begin{align}
s(t)=\sum_{m=0}^{M-1}\sum_{n=0}^{N-1}X_{n,m}e^{j2\pi m \Delta_f (t-nT)} g_{\rm tx}(t-nT),
\end{align}
where $\Delta_f$ is subcarrier spacing and $T = \frac{1}{\Delta_f}$ is the symbol duration.
The above equation can be viewed as a two-step process that first transforms the TF domain symbols to time domain symbols and then shapes the pulse with function $g_{\rm tx}(t)$.
For a linear time-variant channel, the received signal can be expressed as
\begin{equation}\label{TF_model}
r\left( t \right) = \int {\int {h\left( {\tau ,\nu } \right)} } {{e^{j2\pi \nu \left( {t - \tau } \right)}}}s\left( {t - \tau } \right)d\tau d\nu + n(t),
\end{equation}
where the channel impulse response in the DD domain with $P$ resolvable paths is given by\cite{RavitejaOTFS}
\begin{equation}
{h\left( {\tau ,\nu } \right)} = \sum_{i=1}^{P} {h_i} \delta(\tau-\tau_i)\delta(\nu-\nu_i),
\end{equation}
where $\tau_i$ and $\nu_i$ denote the delay and Doppler shift associated with the $i$th path.

At the receiver side, we adopt $g_{\rm rx}^{*}(t)$ as the receiving filter and transform the received signal to the TF domain via performing DFT.
By sampling at $t=nT$ and $f=m\Delta_f$, the received samples $Y_{n,m}$ is given by
\begin{align}\label{TF_Rx}
{Y}_{n,m} = \int {r\left( t \right)g_{{\rm{rx}}}^ * \left( {t - nT} \right){e^{ - j2\pi m\Delta f\left( {t - nT} \right)}}dt}.
\end{align}
According to \cite{RavitejaOTFS}, invoking ideal shaping and receiving pulses $g_{\rm tx}(t)$ and $g_{\rm tx}(t)$ can simplify the TF domain input-output relationship as
\begin{align}\label{TF_in_out}
Y_{n,m} = H_{n,m}X_{n,m} + W_{n,m},
\end{align}
where $H_{n,m}$ is the equivalent channel in the TF domain and $W_{n,m} \sim {\cal CN}(0,\sigma^2)$ is the additive white Gaussian noise with a noise power of $\sigma^2$.
Finally, we arrive at the DD domain received sample $y_{k,l}$ by employing the symplectic finite Fourier transform (SFFT), formulating as
\begin{equation}\label{DD_in_out}
y_{k,l}
= \sum\nolimits_{k' = 0}^{N - 1} {\sum\nolimits_{l' = 0}^{M - 1} {d_{k',l'}} } {h_{k,l}}\left[ {k', l'} \right]+w_{k,l},
\end{equation}
where
\begin{align}
	{h_{k,l}}\left[ {k', l'}\right] &= \sum_{i=1}^{P} {h_i} w\left[k \hspace{-1mm}-\hspace{-1mm} k'\hspace{-1mm}-\hspace{-1mm}k_{\nu_i}, l \hspace{-1mm}-\hspace{-1mm} l'\hspace{-1mm}-\hspace{-1mm}l_{\tau_i}\right] {e^{ - j2\pi \frac{k_{\nu_i}l_{\tau_i}}{NM}}} \;\\\text{and}~
	w\left[k,l\right] &= \frac{1}{NM}\sum\limits_{n = 0}^{N - 1} \sum\limits_{m = 0}^{M - 1} {e^{ - j2\pi nT\frac{k}{NT}}}{e^{j2\pi m\Delta f \frac{l}{M \Delta f}}},
\end{align}
where $k_{\nu_i}$ and $l_{\tau_i}$ denote the Doppler frequency shift index and the delay index associated with the $i$th path, respectively, i.e., $\nu_i = \frac{k_{\nu_i}}{NT}$ and $\tau_i = \frac{l_{\tau_i}}{M\Delta f}$.

\section{Receiver Design}


\subsection{Variational Bayes Approach}
To facilitate the presentation of the proposed variational Bayes approach, we stack the transmitted symbols and equivalent channels into vectors yileding the following compact form of \eqref{DD_in_out}:
\begin{align}\label{DD_in_out_Matrix}
y_{k,l} &= {\tb{h}}^\textrm{T}_{k,l} \tb{d} + w_{k,l},
\end{align}
where $\tb{d}$ and $\tb{h}_{k,l}$ denotes the data symbol vector and equivalent channel vector, respectively, whose the $(k'M+l')$th entry is $d_{k',l'}$ and $h_{k,l}[k',l']$.
As shown in Sec. IV. B of \cite{RavitejaOTFSCE}, using both ideal pulse and practical rectangular pulse provides the same input-output relationship form as in \eqref{DD_in_out_Matrix} by slightly changing the equivalent channels.
Therefore, it is worth to note that the proposed detection in this work is applicable to both cases.
Furthermore, by stacking $y_{k,l}$ into a vector $\tb{y}$, the optimal MAP detection can be formulated as:
\begin{align}\label{OTFS_MAP}
\hat{\tb{d}}=\arg\max_{\tb{d}\in\chi} p(\tb{d}|\tb{y}).
\end{align}
Solving the optimization problem in \eqref{OTFS_MAP} requires a computational complexity order of $|\chi|^{NM}$, which increases exponentially with the size of $\tb{d}$.
As a compromise approach, we focus on the variational Bayes method to handle \eqref{OTFS_MAP} in the sequel.
Specifically, the idea behind variational Bayes aims for finding a distribution ${q}(\tb{d})$ from a tractable distribution family $\mathcal{Q}$ as an optimized approximation of the \emph{a posteriori} distribution $p(\tb{d}|\tb{y})$.
The approximation ${q}(\tb{d})$ can be obtained by minimizing the Kullback-Leibler divergence \cite{kullback1951information} $\mathcal{D}(q||p)$, i.e.,
\begin{align}
{q}^{*}(\tb{d})&=\arg\min_{q\in \mathcal{Q}}\mathcal{D}(q||p)\nonumber\\&=\arg\max_{q\in \mathcal{Q}}\underbrace{\mathbb{E}_{q}\left[-\ln q(\tb{d})+\ln p(\tb{d}|\tb{y})\right]}_\mathcal{L},\label{KLD}
\end{align}
where the expectation is taken over $\tb{d}$ according to the probability density function $q(\tb{d})$.
The functional $\mathcal{L}$ on the right-hand side of \eqref{KLD} is referred to as the evidence lower bound (ELBO), which characterizes the difference of the distribution of latent variables and the distribution of respective observed variables\cite{giordano2018covariances}.
Obviously, the family $\mathcal{Q}$ manages the complexity of the optimization problem in \eqref{KLD}.
In the sequel, we consider a family $\mathcal{Q}$ that all variables in ${q}(\tb{d})$ are mutually independent, satisfying ${q}(\tb{d})=\prod_{k,l} q_{k,l}(d_{k,l})$, also known as mean filed approximation.
Consequently, each latent variable $d_{k,l}$ is characterized by its own variational factor $q_{k,l}(d_{k,l})$.

With the mean field approximation, ${q}(\tb{d})$ can be determined iteratively by maximizing the ELBO.
Since the noise samples $w_{k,l}$ and data symbols $d_{k,l}$ in the DD domain, $\forall k,l$, are independent, $p(\tb{d}|\tb{y})$ can be rewritten as
\begin{align}
p(\tb{d}|\tb{y}) = \prod_{k,l} p(d_{k,l}) p(y_{k,l}|\tb{d}).
\end{align}
Since SFFT is an orthogonal transformation that does not change the statistics of noise samples, $w_{k,l}$ is still Gaussian distributed with zero mean and variance $\sigma^2$, leading to a Gaussian likelihood function $p(y_{k,l}|\tb{d})$.
After some straightforward manipulations, we can rewrite $p(\tb{d}|\tb{y})$ in a pairwise form, i.e.,
\begin{equation}\label{post}
p(\tb{d}|\tb{y}) \propto \prod_{k,l} \zeta_{k,l}(d_{k,l}) \prod_{k,l} \psi_{k,l} (d_{k,l},d_{k',l'}), \vspace{-4mm}
\end{equation}
where
\begin{align}
	\zeta_{k,l}(d_{k,l})=p(d_{k,l}) \exp\left(-\frac{\rho_{k,l}|d_{k,l}|^2+\epsilon_{k,l} d_{k,l}}{\sigma^2}\right)\; \\\text{and}~
	\psi_{k,l} (d_{k,l},d_{k',l'})=\exp\left(-\frac{\varrho_{k,l,k',l'} d_{k,l} d_{k',l'}}{\sigma^2}\right),
\end{align}
with $\rho_{k,l}=\sum_{k',l'} |{h}_{k',l'}[k,l]|^2$, $\epsilon_{k,l}=2\sum_{k',l'} \mathcal{R}\{{h}_{k',l'}[k,l]\cdot y_{k'.l'}\}$, and $\varrho_{k,l,k',l'}=2\mathcal{R}\{{h}_{k,l}[k,l]{h}^{*}_{k,l}[k',l']\}$.
Having $p(\tb{d}|\tb{y})$ in hand, we substitute \eqref{post} and $q(\tb{d})$ into $\mathcal{L}$ which yields
\begin{align}
&\mathcal{L}=\mathbb{E}_{q}\left[\sum_{{k,l}} \ln \psi_{k,l}  (d_{k,l},d_{k',l'})-\sum_{k,l}\ln\frac{q_{k,l}(d_{k,l})}{\zeta_{k,l}(d_{k,l})}\right]\nonumber\\
&=\mathbb{E}_{q}\Bigg[-\frac{\sum_{k,l}\varrho_{k,l,k',l'} d_{k,l} d_{k',l'}}{\sigma^2}-\sum_{k,l}\ln\frac{q_{k,l}(d_{k,l})}{\zeta_{k,l}(d_{k,l})}\Bigg].
\end{align}
Next, we aim for finding a stationary point of the variational $\mathcal{L}$, which is given by the solution of $\partial\mathcal{L}/\partial q(\tb{d})=0$.
According to Euler-Lagrange equation\cite{gelfand2000calculus}, this requires the partial derivatives of $\mathcal{L}$ with respect to all local functions $q_{k,l}(d_{k,l}),~\forall k,l$, being zero.
To this end, we propose a simple iterative algorithm to update each local function.
Let us take the latent variable $d_{k,l}$ in the $\mathrm{iter}$-th iteration as an example to state the proposed algorithm.
Given the obtained local functions in the $\left(\mathrm{iter}-1\right)$-th iteration $q^{\mathrm{iter}-1}_{k',l'}(d_{k',l'})$, $\forall {\{k',l'\}\neq \{k,l\}}$, setting the partial derivative $\partial \mathcal{L}/\partial q_{k,l}(d_{k,l})$ to zero leads to
\begin{align}\label{EqL}
\mathbb{E}_{q\backslash{k,l}} &\left[-\frac{1}{\sigma^2}\!\!\sum_{k',l'}\varrho_{k,l,k',l'} d_{k,l} d_{k',l'}\right]\nonumber\\&+\ln \zeta_{k,l}({d}_{k,l}) -\ln q^{\mathrm{iter}}_{k,l}(d_{k,l}) +C=0,
\end{align}
where ${q}\backslash\{k,l\}=\prod_{\{k',l'\}\neq \{k,l\}}{q^{\mathrm{iter}-1}_{k',l'}(d_{k,l})}$ and $C$ denotes a constant.
%
%
Then, solving \eqref{EqL} for $q_{k,l}(d_{k,l})$ results in the local distribution, expressed as
\begin{align}\label{qfunction}
  q_{k,l}^{\mathrm{iter}}(d_{k,l})\!\!\propto& \zeta_{k,l}({d}_{k,l}) \!\exp\!\!\left(\!\mathbb{E}_{{q}\backslash\{k,l\}} \!\!\left[\!-\frac{1}{\sigma^2}\!\sum_{k',l'}\!\varrho_{k,l,k',l'} d_{k,l} d_{k',l'}\!\right]\!\right)\nonumber\\
  \propto&p({d}_{k,l})\exp\left(-\frac{\rho_{k,l}|d_{k,l}|^2-m_{k,l} d_{k,l}}{\sigma^2}\right),
\end{align}
where $m_{k,l}=\epsilon_{k,l}-\sum_{\{k',l'\}\neq \{k,l\}}\varrho_{k,l,k',l'}\mathbb{E}_{q^{\mathrm{iter}-1}_{k',l'}}[d_{k',l'}]$.
%
In a similar way, we repeat the above procedure to approximate \emph{a posteriori} distributions for all the data symbols iteratively, resulting in the approximate marginals $q_{k,l}^{*}(d_{k,l})$, $\forall k,l$.
Then, the sophisticated MAP detection in \eqref{OTFS_MAP} is simplified to a problem of finding the maximum of marginal distribution $q_{k,l}^{*}(d_{k,l})$, i.e.,
\begin{align}
\hat{d}_{k,l}=\arg\max_{d_{k,l}\in\chi} q^{*}_{k,l}(d_{k,l}).
\end{align}
It is easy to see that the complexity of the detection is reduced to the order of $|\chi| NM$, which is significantly lower than that of the conventional MAP detection.

\subsection{Comparison with MP Receiver}
It is well known that the MP receiver converges to local optimum if the factor graph has loops, especially in a rich-scattering environments.
Actually, even when the number of paths is as small as $2$, there may exist a large number of girth-4 loops. 
%
However, for the proposed variational Bayes approach, it can be observed that the second order partial derivative of $\mathcal{L}$ with respect to $q_{k,l}(d_{k,l}),\forall k,l$, is
\begin{align}
\frac{\partial^2 \mathcal{L}}{\partial q_{k,l}(d_{k,l})^2}=-\frac{1}{q_{k,l}(d_{k,l})}\leq 0,
\end{align}
while $\frac{\partial^2 \mathcal{L}}{\partial q_{k,l}(d_{k,l}) \partial q_{k',l'}(d_{k,l})}=0$.
Hence, the Hessian matrix of the ELBO is diagonal with the $(kM+l)$th diagonal entry being $-\frac{1}{q_{k,l}(d_{k,l})}$.
This observation indicates that $\mathcal{L}$ is a concave functional and its local maximum is the global one.
Therefore, iteratively updating each local distribution ${q_{k,l}(d_{k,l})}$ according to \eqref{qfunction} is guaranteed to converge to an optimal approximate distribution $q^{*}(\tb{d})$ that maximizes the ELBO.

Moreover, in terms of computational complexity, both the MP receiver and the proposed detector achieve a linear complexity with the number of symbols.
Nevertheless, we can observe from \eqref{qfunction} that the update of $q_{k,l}(d_{k,l})$ depends on the marginal means of all the other symbols, which implies that the marginal means need to be  calculated only once in a single iteration.
Therefore, for a time-varying channel consisting of $P$ propagation paths, the total complexity of the proposed algorithm is $\mathcal{O}(|\chi|NMP)$.
On the contrary, for the MP receivers, the messages have to be calculated individually for different connected factor vertices, leading to a computational complexity of $\mathcal{O}(|\chi|NMP^2)$.

To sum up, the proposed variational Bayes approach is attractive in rich-scattering environments, owing to its low-complexity and guaranteed convergence feature.

\section{Numerical Results}
This section illustrates the numerical results via Monte Carlo simulations.
All simulation results are averaged from $2\times 10^4$ OTFS frames and for each OTFS frame, we set $M=512$ and $N=128$ indicating that there are $128$ time slots and 512 subcarriers in the TF domain.
The carrier frequency is $4$ GHz and the subcarrier spacing is $15$ kHz.
Quadrature phase shift keying (QPSK) modulation is utilized for symbol mapping.
The speed of the mobile user is set to be $v=120$ km/h, leading to a maximum Doppler frequency shift index $k_{\nu_{\rm max}}=4$.
We assume that the channel information is perfectly known at the receiver and the maximum delay index is $l_{\tau_{\rm max}}=10$.
The Doppler index of the $i$th path is drawn with equal probabilities from the set $[-k_{\nu_{\rm max}},k_{\nu_{\rm max}}]$ and the delay index belongs to $[1,l_{\tau_{\rm max}}]$ excluding the first path which always satisfies $l_{\tau_1}=0$.
The channel coefficients $h_i$ are generated according to the distribution $h_{i}\sim \mathcal{CN}(0,q^{l_{\tau_i}})$, where the normalized power delay profile $q^{l_{\tau_i}}=\frac{\exp(-0.1 l_{\tau_i})}{\sum_i \exp(-0.1 l_{\tau_i})}$.
The number of iterations is set to $10$.

\begin{figure}[t]
\vspace{-3mm}
\centering
\includegraphics[width=3.5in]{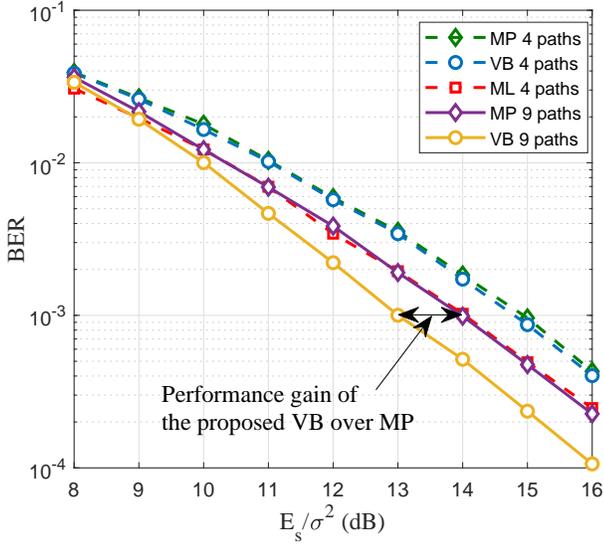}\vspace{-3mm}
\caption{Impact of number of paths on the BER performance.}%
\label{paths}%
\vspace{-3mm}
\end{figure}

We compare the bit error rate (BER) performance versus ${\rm{E}_{\rm{s}}}/\sigma^2$ corresponding to the proposed variational Bayes (VB) method and the conventional MP for OTFS modulation over delay-Doppler channels in Fig. \ref{paths}, where ${\rm{E}_{\rm{s}}}$ denotes the symbol energy.
Two scenarios with $P=4$ and $P=9$ paths are considered.
It is observed that in the 4-path scenario, the proposed algorithm and the MP method have virtually the same performance.
When $P=9$, the performance of the proposed VB method is significantly better than that of the MP receiver.
This is because in the $P=9$ scenario, the MP receiver converges to a local optimum due to the large amount of girth-4 loops.
Moreover, the performance improvement brought by our proposed VB detector is magnified by the increasing number of paths, compared to the MP receiver.
In fact, due to its convergence-guaranteed property, the proposed receiver can exploit the TF diversity gain more efficiently offered by OTFS modulation than that of the MP receiver.
As a benchmark algorithm, the performance for the MAP detector with $4$ paths is illustrated\footnote{The MAP detection is obtained by exhaustive search. Note that, the curve for the MAP detector with $9$ paths is not shown due to the prohibitively high computational complexity.}.
We can observe that the proposed VB-based receiver approaches the performance of the MAP detector while the complexity is significantly reduced.

\begin{figure}[t]
 \vspace{-3mm}
\centering
\includegraphics[width=3.5in]{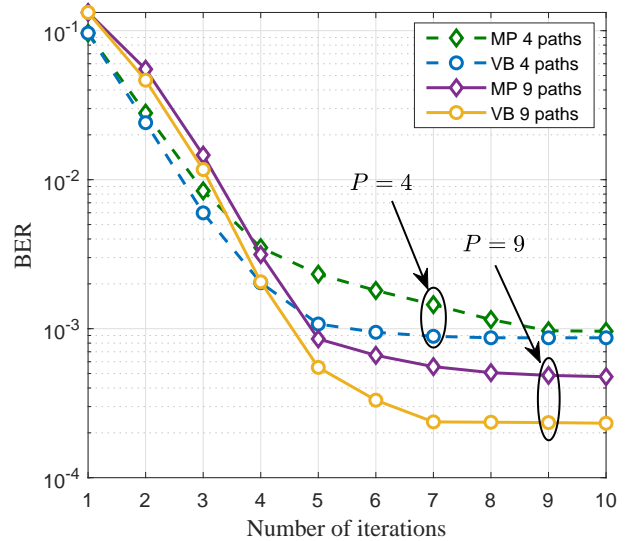}\vspace{-3mm}
\caption{BER performance versus the number of iterations.}
\label{iterations}
\vspace{-3mm}
\end{figure}

To illustrate the convergence behaviors, we plot the BER performance of the proposed VB method and the MP method versus the number of iterations in Fig. \ref{iterations}, where ${\rm{E}_{\rm{s}}}/\sigma^2$ is set to $15$ dB.
Again, two scenarios with different number of paths are included.
We see that increasing the number of iterations leads to a better performance for both algorithms.
Moreover, benefiting from the convexity of the ELBO, the proposed VB algorithm converges much faster than the MP receiver in the 4-path scenario even they converge to almost the same BER.
On the other hand, for the 9-path scenario, the proposed VB algorithm can converge faster and achieve a better detection performance compared to that of the MP receiver.
Thus, the proposed VB method is very attractive in OTFS systems, especially when the number of paths is large.


\section{Conclusions}
This paper proposed a variational Bayes-based receiver for the promising OTFS modulation technology.
To reduce the complexity of the conventional MAP receiver, we approximate the joint \emph{a posteriori} distribution by the product of local marginals via maximizing the ELBO.
By choosing a distribution family for the VB method appropriately, the ELBO is convexified which guarantees the convergence of the proposed variational Bayes approach.
We further analyzed the complexity of the proposed algorithm and showed that it is lower than that of both MAP and MP receiver.
Simulation results confirmed the fast convergence of the proposed algorithm and its superior detection performance compared to the MP algorithm, especially in practical multipath scenarios.
\bibliographystyle{IEEEtran}
\bibliography{otfs}
\end{document}